\documentclass[journal]{vgtc}                     

\onlineid{1008}

\vgtccategory{Research}

\title{Investigating Ethical Data Communication with \textit{Purrsuasion}: \\ An Educational Game about Negotiated Data Disclosure}

\author{Krisha Mehta, Sami Elahi, and Alex Kale }

\authorfooter{
  \item
  	Krisha Mehta is with the University of Chicago.
  	E-mail: krisha@uchicago.edu
  \item
  	Sami Elahi is with the University of Chicago.
  	E-mail: melahi@uchicago.edu

  \item Alex Kale is with the University of Chicago.
  	E-mail: kalea@uchicago.edu.
}
 
\abstract{
Data communication entails ethical dilemmas where situational constraints forbid full disclosure of source data. Whereas visualization research and pedagogy often frames ethics as a matter of individuals making deceptive design choices or being misled, disclosure problems involve negotiation between pro-social actors. To provide observability into these situated judgments, we contribute \textit{Purrsuasion}, an open-source visualization game where participants play the roles of (i) data providers designing visualizations subject to disclosure constraints and (ii) data seekers requesting information and awarding a contract. We deploy \textit{Purrsuasion} in an undergraduate data science class (N = 27), gathering gameplay data to support a mixed-methods analysis of students’ communication dynamics, problem solving, and trust formation. We find that difficulties envisioning an ideal visualization solution lead to satisficing in visualization authoring and difficulties attributing authorial intent. Given these challenges, we approach scoring student solutions by developing a heuristic rubric that supports sociotechnical judgments of disclosure adherence.
}


\teaser{
  \centering
  \includegraphics[width=\linewidth, alt={A view of a city with buildings peeking out of the clouds.}]{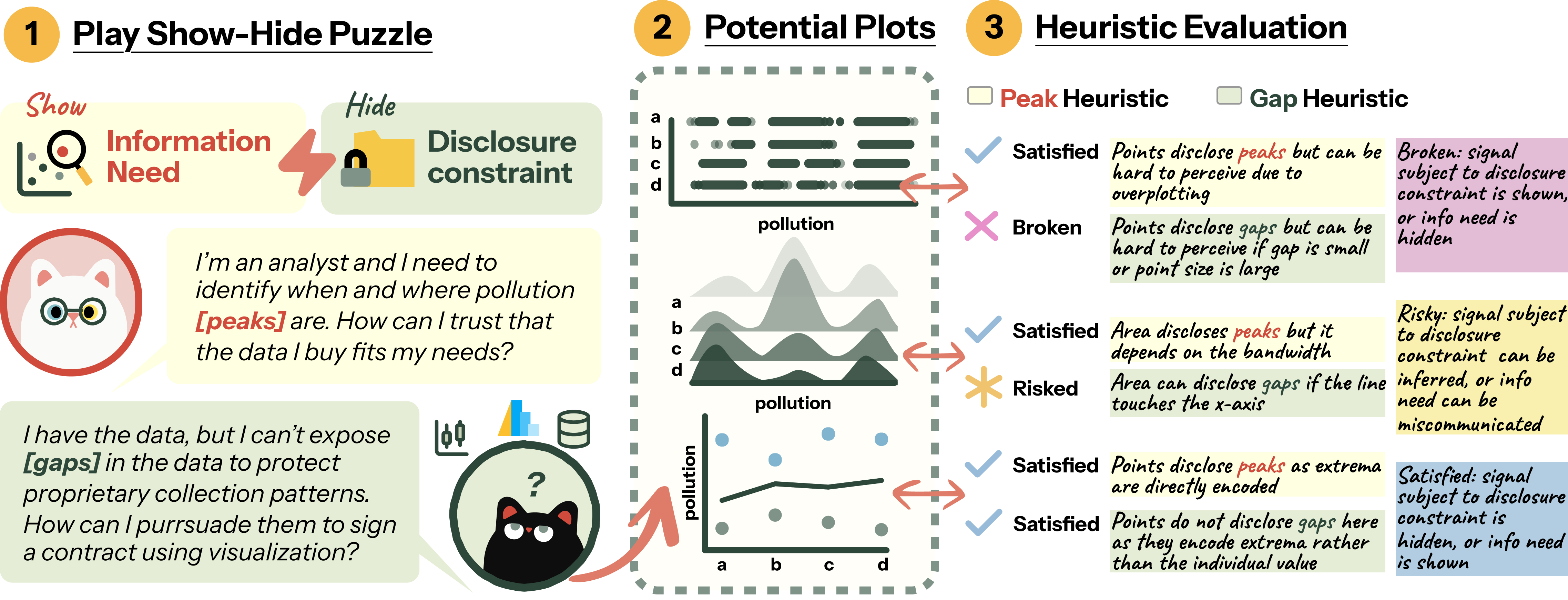}
  \caption{In \textit{Purrsuasion}, students (1) navigate a set of data communication scenarios represented by show-hide puzzles, and (2) ideate potential design solutions in an attempt to balance an information need with a disclosure constraint. (3) Instructors use a heuristic rubric to evaluate the visualization's success in balancing the conflicting, situated requirements of the puzzle.}
  \label{fig:teaser}
}

\graphicspath{{figs/}{figures/}{pictures/}{images/}{./}} 

\usepackage{tabu}                      
\usepackage{booktabs}                  
\usepackage{lipsum}                    
\usepackage{mwe}                       
\usepackage{ccicons}                   

\usepackage{mathptmx} 
 \usepackage{amsmath}
\usepackage{booktabs}
\usepackage{tabularx}
\usepackage{array}
\usepackage{enumitem}
\usepackage{xurl}
\usepackage{soul}
\usepackage{xcolor}
\usepackage[table]{xcolor}
\usepackage{dblfloatfix} 
\usepackage{pdfpages}

\newcolumntype{Y}{>{\raggedright\arraybackslash}X}

\newcommand{\schemacell}[1]{%
  \begin{minipage}[t]{\linewidth}\vspace{0pt}\small
    \begin{itemize}[leftmargin=1.1em,labelsep=0.4em,nosep]
      #1
    \end{itemize}
  \end{minipage}%
}

\definecolor{quoteColor}{HTML}{B3DDF2}

\newcommand{\pxx}[1]{\textbf{S$_{#1}$}}

\definecolor{definitionColor}{HTML}{E1EBD1}

\newcommand{\eg}{{e.g., }}

\begin{document}


\firstsection{Introduction}

\maketitle


Visualization research and pedagogy treats ethical data communication as a central concern.
Numerous studies investigate the impact on human data interpretation when visualizations incorporate potentially misleading design patterns---\eg truncating the y-axis~\cite{correll2020-yaxis, long2024-cutYaxis}, violating user expectations based on graphical conventions~\cite{Zacks1999, Kerns2021, tufte1983}, or distorting the correspondence between data and visual encoding channels~\cite{correll2017-blackhat,kindlmann2014-AVD, McNutt2020-Mirages}.
Work on visualization literacy develops assessments of a chart user's ability to avoid being misled by 
such design patterns
(\eg~\cite{Ge2023-Calvi, lee2016vlat}).
Visualization pedagogy tends to emphasize identifying and avoiding specific deceptive practices as a learning objective, with coursework often asking students to either find or create examples that commit these violations as an exercise to prompt reflection.
In this way, both research and teaching rely on conceptualizations of deceptive visualization that focus on the cognition and design processes of individuals, rather than situating ethical visualization 
in the communication dynamics of two or more people.
We argue that 
this focus
on individual judgment risks 
fixating on
misuse of visualization as a moral hazard, while underexamining legitimate 
use cases that
require designers to avoid full transparency with their audience.

In practice, visualization designers encounter morally gray scenarios where tensions arise between their responsibilities to different parties.
For example, an environmental analyst may need to communicate pollution peaks while masking proprietary sensor locations. (Fig.~\ref{fig:teaser}).
The environmental analyst is not a malicious actor seeking to deceive, but a pro-social one puzzling out a complex design trade-off.
Following prior work, we characterize such design scenarios as problems of \textbf{\textit{data disclosure}}~\cite{mehta2025designingDisclosure}, which require designers to balance showing information that is useful to their audience and hiding information they are constrained not to reveal.
We define design goals for data disclosure problems in terms of the specific \textbf{\textit{data signals}} (\eg outliers, clusters, gaps) or aspects of a dataset that are the subject an audience's information need or a designer's data sharing constraint.
\looseness=-1

Design scenarios requiring selective data disclosure present critical unmet challenges for the visualization community.
Prior work demonstrates that existing visualization tools are not built to optimize for design goals around disclosure~\cite{Wu2024-transforms}, leaving visualization designers with a ``guess and check'' workflow as their only recourse to find suitable designs in such scenarios~\cite{mehta2025designingDisclosure}.
Developing the judgment to navigate design trade-offs around responsible data disclosure is a critical learning outcome for data science students, yet to our knowledge it is currently missing from most data visualization curriculum.
Finally, the social dynamics that surround these design scenarios---potentially motivating the behavior of a designer or the trust of their audience---lack salience in research and teaching materials.
In this study, we investigate these ``entangled''~\cite{Akbaba2025entanglements} challenges by developing a novel game platform for research and teaching, which enables us to explore problem solving and moral reasoning around selective data disclosure with our students.

We present \textit{Purrsuasion}, a cat-themed browser-based game that asks participants solve data disclosure problems through visualization authoring and negotiation.
Players are assigned the roles of (i) a \textit{receiver} or data seeker who makes a request for information and (ii) a \textit{sender} or data provider who authors visualizations subject to a disclosure constraint.
Each round of gameplay centers on a distinct \textbf{\textit{show-hide puzzle}}: a design scenario where the receiver is endowed with an information need or data signal they wish to learn about, and the sender is endowed with a disclosure constraint or a data signal they are forbidden to reveal (Fig.~\ref{fig:teaser}).
To enable scoring of visualizations as solutions to these puzzles, we define a heuristic rubric to help instructors and researchers assess disclosure adherence.
We develop \textit{Purrsuasion}, show-hide puzzles, and the heuristic rubric to encapsulate disclosure problems and their inherent ethical dilemmas in a controlled setting, so we can provide experiential learning for students while also gaining observability into their problem solving, communication dynamics, and trust formation in such scenarios.
Presenting the interface as a game provides structure and social permission for collaboration and reflection~\cite{brandt2004_collabThroughGames}.
We contribute an open-source implementation of the \textit{Purrsuasion} platform as a resource for the community, including onboarding materials and instructor tools for running, scoring, and adapting the game.

We also contribute a mixed-methods study based on a deployment of \textit{Purrsuasion} in a visualization course for data science undergraduates at [institution redacted].
$27/33$ students participating in the game consent to donate their interaction log data resulting from gameplay and fill out a survey reflecting on the ethics of data disclosure.
In addition to characterizing communication dynamics, we attend in particular to how students conceptualize solutions to disclosure problems.
To support this analysis, we draw on Mehta et al.'s \textbf{\textit{taxonomy of disclosure tactics}}~\cite{mehta2025designingDisclosure}, which provides a deductive framework for describing how design choices about data transformations determine what information is revealed in a visualization (see Section~\ref{analysis_method}).
Our analysis shows that students face a \textbf{\textit{``gulf of envisioning'' in visualization}}~\cite{subramonyam2024_gulfofEnvision}, a fundamental difficulty conceiving of an optimal design to satisfy design goals around disclosure.
We describe how this gulf precipitates design fixation in visualization authoring for senders, difficulties attributing authorial intent for receivers, and pitfalls in negotiated data sharing as a result.
Through the development of the rubric, we discover that evaluating student solutions requires sociotechnical judgments of situated communication risks, subverting our expectation that disclosure adherence could be evaluated through automated constraint checking.
Taken together, these findings suggest that ethical data disclosure through visualization requires more interactive support for ideation, interpretion, query formulation, and even trust repair in instances of miscommunication.
We discuss how our game platform can be adapted to support future research---\eg through the use of different disclosure puzzles to encapsulate design scenarios such as forecasting for a decision-maker or data fusion from multiple sources of varying data quality.

\section{Background}
We contextualize \textit{Purrsuasion} by highlighting how it offers a new perspective on ethical visualization and visualization literacy, and how it extends a long standing tradition of educational games as a way to engage students and surface what they learn through play.

\subsection{Ethical Visualization and Visualization Literacy}
A growing body of 
research emphasizes that ethics is inseparable from a visualization's communicative power and its consequences for how people reason, decide, and form opinions~\cite{correll2019-ethics}. While visualization ethics spans affect, values, and politics in addition to accuracy and responsibility~\cite{ehmel2021topographyViolence, Lee-Robbins2023-affective, riche2018dataDrivenStorytelling, dork2013-criticalVis}, much of the visualization research and pedagogy centers on design decisions that 
produce
misleading visualizations that can harm or deceive viewers~\cite{correll2017-blackhat,lo2022misinformed, correll2020-yaxis, long2024-cutYaxis, McNutt2020-Mirages}. While learning to recognize and avoid deceptive design is essential, narrowing ethics to 
a focus on
deception alone underrepresents the kinds of trade-offs that arise in real-world data communication, especially in \textit{two ways}. 
First, a deception-centered framing can overstate the prevalence of intentionally malicious designers---\eg relative to a student's future professional experiences, which we argue might entail responsibilities to avoid full transparency about sensitive data sources (see Section 1). 
We aim to understand what kinds of design practices visualization students use and consider permissible in these design scenarios.
To do so, we draw on Mehta et al.'s \textbf{\textit{taxonomy of disclosure tactics}}~\cite{mehta2025designingDisclosure} to categorize the design operations that directly influence what data signals are exposed by a chart.
Our study presents the first use of this framework in an empirical investigation to evaluate designers' and audiences' ethical reasoning about data disclosure through visualization.

Second, the ethical significance of design is fundamentally relational. It emerges in the interaction between a designer’s choices and an audience’s incentives, knowledge, and interpretation. 
Research that centers both sides of the designer-audience dyad remains limited, particularly work that studies how designers and audiences jointly navigate ethical tensions in an interactive setting~\cite{zhang2024-sharedInfoDisplays,nanayakkara2022-privacyUtility}. Our ways of measuring and teaching visualization literacy often incorporate ethics through a deception-focused, individual-level lens~\cite{Camba2022-DeceptionVisLit}. Common
visualization literacy assessments
such as VLAT~\cite{lee2016vlat}, Mini VLAT~\cite{Pandey2023-miniVLAT}, and CALVI~\cite{Ge2023-Calvi} assess whether a viewer can extract information from common chart types, treating visualization as an isolated object to be decoded. In this framing, the potential for deception
is often reduced to an individual’s ability to spot misleading graphics and critique a finished artifact. Other works such as AVEC~\cite{ge2025-avec} shift toward visualization construction, but still evaluate individuals in isolation rather than the designer-audience relationship. 
\textit{Purrsuasion} makes a dyadic view of ethics and literacy observable by placing students in designer and audience roles where they must negotiate what counts as sufficient, persuasive, and compliant visual evidence under disclosure constraints. 

\subsection{Educational Games as Pedagogy and Probe}

HCI has a long-standing tradition of using educational games and simulations to make complex concepts both teachable and learnable\cite{Ye2025_awarenessToAction, Malone1981_theoryOfInstrinsicMotivation, Bai2022_gamificationImproveLearning}. Building on this tradition, communication games \cite{Riegelsberger2003_researchersDilemma, FAIR2022_Diplomacy} offer HCI researchers a way to study how people reason under structured roles, partial information, and competing goals. Visualization research therefore stands to benefit from studying communication games: interactions in which participants make strategic choices about how and what to communicate through visualizations. Game-based approaches have already been used productively in visualization contexts \cite{adar2023roboviz, Adelberger2025_Iguanodon}. We draw on this tradition by presenting a game inspired in part by negotiation activities common in business school programs. Observing disclosure through a game is methodologically useful because it turns otherwise tacit design trade-offs into visible, consequential actions that can be compared across players, rounds, and roles. In \textit{Purrsuasion}, students make decisions about disclosure and presentation while designing  
to show the data signals required
for common analysis tasks ~\cite{amar2005low, correll2019-LooksGoodToMe, kim2018assessing, sarikaya2017scatterplots, shneiderman2003eyes, wilkinson2008scagnostics}. 
By developing and deploying the game,
we contribute both empirical evidence about visualization-mediated communication and a classroom-ready activity that surfaces the communicative and ethical dimensions of visualization design.

\section{\textit{Purrsuasion}: The Game}
\label{the_game}
We designed \textit{Purrsuasion} to model ethical dilemmas in real-world data communication around data sharing, where the data seeker must judge a dataset’s usefulness as the data provider is constrained in what they can disclose. 
Such a dynamic shows up in many contexts where the data provider cannot reveal sensitive information upfront, whether due to privacy, proprietary, or policy constraints.
\textit{Purrsuasion} studies this ethical dilemma 
through
a simplified data marketplace 
scenario.
We chose this framing because negotiating a data sharing agreement provides a familiar, minimal structure for explaining the game mechanics.


\subsection{How to Play the Game}
The game is played in groups of three and consists of two roles: data providers that we dub \textbf{senders} and data seekers that we dub \textbf{receivers}. 
Each round of the game has two senders and one receiver.
The receiver's goal is to obtain visual evidence that addresses a given \textit{information need} (\eg identify peaks in pollution data in Fig.~\ref{fig:teaser}). Both senders have access to the same dataset, which they use 
to produce a visualization and explanation to satisfy the receiver's information need while respecting a \textit{disclosure constraint} that limits what can be revealed (\eg hide gaps to protect proprietary data collection patterns in Fig.~\ref{fig:teaser}). After reviewing the sender responses, the receiver selects one sender as the winner of the round and signs a hypothetical contract with them to gain access to the full dataset. The game consists of three rounds, where each round introduces a different dataset, information need, and disclosure constraint. 
\looseness=-1

\begin{figure}[t]
    \centering
    \includegraphics[width=\columnwidth]{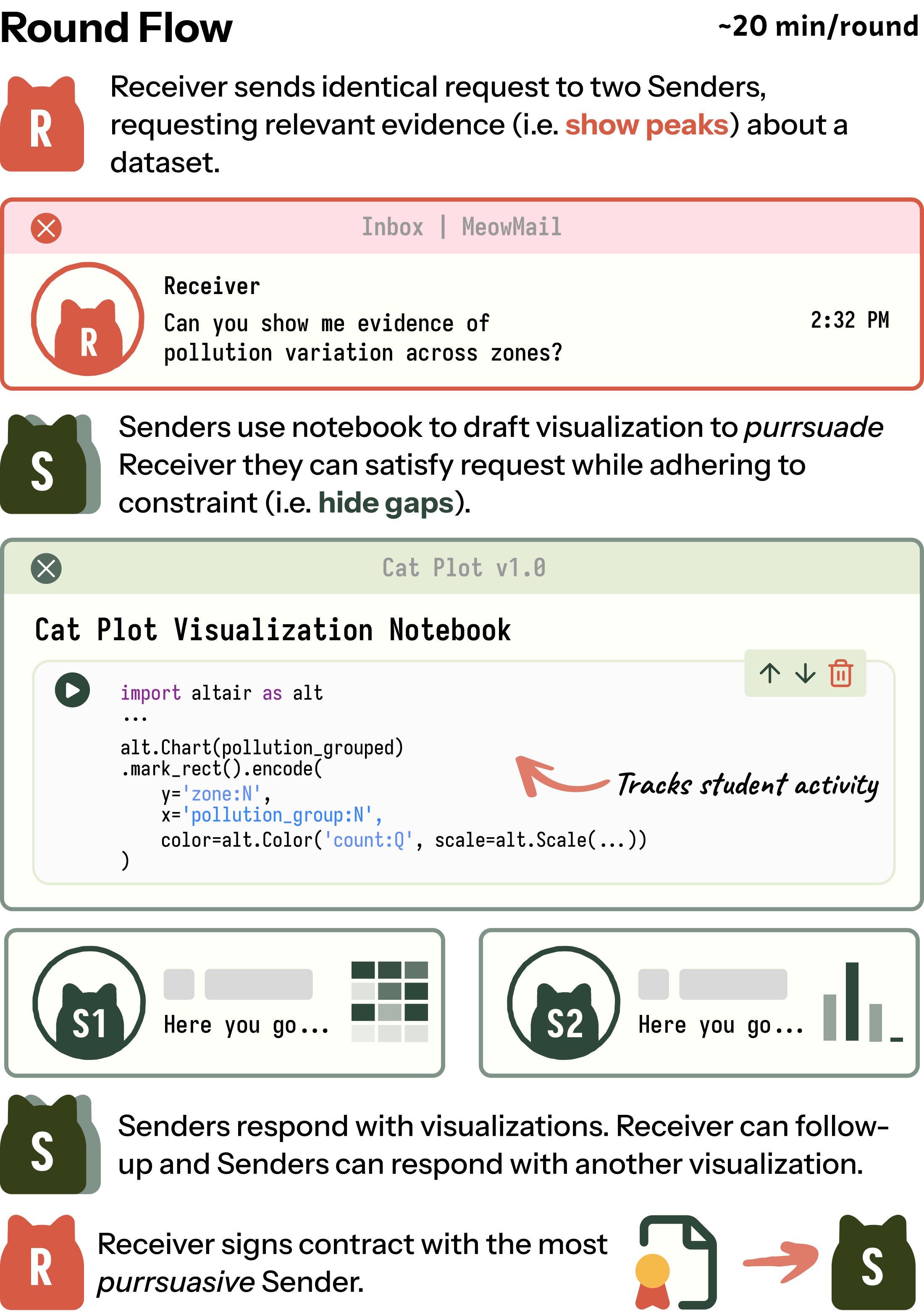}
    \caption{
  A trio of students play a round of \textit{Purrsuasion}.
    }
    \label{fig:game_flow}
\end{figure}

Each round follows a consistent communication sequence described in Figure~\ref{fig:game_flow}. Receivers start each round by making identical queries to both senders. Senders respond with visualizations that answers the receiver's query while adhering to the disclosure constraint. 
Receivers ask a follow-up question to each sender separately, requesting clarification or additional detail, and senders respond with new or updated visualizations. Although the game can support more than two exchanges, our initial implementation limits each sender-receiver interaction to two responses 
to provide senders enough time for design ideation.
The round concludes when the receiver selects one of the two senders to win the contract and provides a short written rationale for their choice.  

\subsection{Game Interface}
\label{interface}

We build \textit{Purrsuasion} as a browser-based cat-themed game, stemming from a pun on the game's ``purrsuasive'' objectives. Upon login, participants enter a fictional operating system called \textit{Mewnix OS}, which provides role-specific applications: \textit{Cat Plot}, \textit{Meow Mail}, and \textit{Whisker Sign}. Senders utilize \textit{Cat Plot} (Fig.~\ref{fig:game_flow}), an interactive notebook environment built using Pyodide, to process data and design visualizations. The application pre-loads the specific datasets required for each round along with essential dependencies, including Altair, Pandas, and NumPy. Communication between players occurs through \textit{Meow Mail} (Fig.~\ref{fig:game_flow}), a WebSocket-based email client designed for the exchange of messages and visualizations. To conclude a round, receivers use \textit{Whisker Sign} to award a contract to the sender who best satisfies their information need. Instructors manage gameplay through an administrator dashboard that allows for roster management, progress monitoring, and the export of anonymized gameplay logs. The dashboard also serves as the operational hub for the scoring rubric detailed in Section~\ref{scoring_solutions}. 
While the interface utilizes whimsical cat-theming to support student buy-in, the underlying design 
codifies a unified data communication
workflow to facilitate observability of
student problem solving.

We view \textit{Purrsuasion} not merely as a standalone game, but as a particular instantiation within a broader data communication puzzle platform (see Section~\ref{extending_the_game}). Following the design philosophy of frameworks such as ReVISit~\cite{cutler2026-revisit},  
we prioritized low-maintenance deployment, extensibility, and instrumentation. We developed \textit{Purrsuasion} using TypeScript across both the front and back ends and selected SQLite as the database such that the application could easily be self-hosted on a single machine. All narrative assets are managed via a global spritesheet and configuration files, allowing instructors to customize the game without modifying the underlying codebase. We release the source code and deployment documentation as an open-source contribution, enabling the visualization community to extend the library of puzzles and utilize the platform for both future experimentation and learning experiences. See \url{https://github.com/anon-vis/purrsuasion}. 

\begin{table*}[!t]
\centering
\caption{Types of puzzles and their corresponding sender-receiver instructions used in the study.}
\small
\setlength{\tabcolsep}{3pt}
\renewcommand{\arraystretch}{1.15}

\begin{tabularx}{\textwidth}{@{} p{2.5cm} Y Y >{\raggedright\arraybackslash}p{5.5cm} @{}}
\toprule
\textbf{Puzzle Type} & \textbf{Receiver Prompt} & \textbf{Sender Prompt} \\
\midrule
\textbf{Show High Saturation, Hide Specific Locations} &
You’re an analyst working with a tenants’ rights group. Your job is to explore how retail stores are spread across the U.S. to understand where saturation is high or low. Your task is to choose a data broker (sender) who you trust to provide the data needed to understand these patterns. &
You’re a data broker who maintains a detailed dataset of retail stores, including their locations and counts across different areas. Your job is to share visualizations showing the dataset contains what your client (the receiver) needs. However, you should not show very fine-grained details (e.g., individual blocks or exact storefront locations) that could enable landlords or large chains to identify stores and raise rents in specific areas. 
\\

\addlinespace[2pt]

\textbf{Show Peaks, Hide Gaps} &
You’re an environmental analyst helping identify when and where pollution levels peak or drop significantly. Your task is to choose a data broker (sender) who you trust to provide the data your team needs to guide policy decisions concerning pollution sources or behaviors. &
You’re a data broker selling a high-resolution air-quality dataset. You want to build trust by demonstrating that the dataset contains the information your potential client (the receiver) needs. But to protect proprietary collection patterns, you cannot expose the exact location of gaps in the data distribution in your visualizations. 
\\
\textbf{Show Outliers, Hide Individual Points} &
You’re a city logistics planner allocating inspection teams for next month. Outliers in either workload or lateness could signal bottlenecks or failing warehouses. You need to identify which warehouses are atypical so you can plan targeted audits. Your task is to choose a data broker (sender) who you trust to provide this information in full upon signing a contract. &
You’re a data broker selling a dataset on warehouse performance. You want to build trust by demonstrating that the dataset contains the information your potential client (the receiver) needs. However, to protect supplier relationships and avoid pinpointing individual warehouses, you should hide warehouse and zone identities in your visualizations.
\\
\bottomrule
\end{tabularx}
\label{tab:puzzle_promptsOnly}
\vspace{-2em}
\end{table*}

\subsection{Data Signals and Show-Hide Puzzles}
\label{showHide}
Each round of the game asks the sender to navigate design trade-offs around showing and hiding different \textbf{\textit{data signals}}: facts or relationships in a dataset that are the subject of the receiver's information need or the sender's disclosure constraint. We call the pairing of data signals, expressed in a receiver's information need and the sender's
disclosure constraint, a \textbf{\textit{show-hide puzzle}}. In order for a show-hide puzzle to present an appropriate challenge, showing the data signal that the receiver needs should be in tension with hiding the data signal the sender is forbidden from revealing. The puzzle will be trivially easy if the two data signals are orthogonal. Otherwise, if the sender's constraint is adversarial to the receiver's need, the puzzle can be impossible to solve or edge into overt deception, which would contravene our learning and research objectives. The puzzles included in this game therefore span a spectrum from nearly orthogonal signal pairs to strongly overlapping ones, creating varying degrees of design tension. 

To generate 
instructions
and datasets for the three puzzles used in our study, we followed an iterative process combining manual effort and ChatGPT (GPT-5) output. We gave both senders access to the same dataset so that differences in the receiver’s contract decision could be attributed to how the data were visualized.
See Supplemental Material for information on stimulus generation, and Section~\ref{sec:appendices} for considerations on generalizing and scaling the puzzle creation process. 
Table~\ref{tab:puzzle_promptsOnly} describes the puzzle types that we used in the game along with the corresponding sender and receiver instructions.
Below, we describe each puzzle, ordered from most to least orthogonal signal pair, giving examples of successful solutions vs. those that violate the disclosure constraint.
\looseness=-1

\textbf{1. Show High Saturation, Hide Specific Locations:} 
The sender must show where store saturation is high without revealing the exact locations of individual stores.
This puzzle illustrates the modifiable areal unit problem (MAUP): apparent “hot spots” can shift when the same data are aggregated into different spatial units (counties vs states vs. regions), making the consequences of choosing a particular level of aggregation directly observable. This puzzle allows us to study not only whether senders satisfy the constraint, but also how they reason about scale, justify design choices, and communicate the limits of the resulting inference.
A strong solution compares multiple aggregation levels---\eg presented in small multiples or a sequence of messages---to show which patterns persist and which are artifacts of the chosen areal unit. A clear constraint violation plots stores as points or otherwise exposes precise locations as shown in Fig.~\ref{fig:gallery}A. 


\textbf{2. Show Peaks, Hide Gaps:}
The sender must communicate where pollution levels peak while concealing where measurements are missing, since gaps can reveal proprietary data collection patterns (Fig.~\ref{fig:teaser}). The design tension is that many natural ways to show peaks such as scatterplots, histograms, 
and densities can
make negative space legible, effectively disclosing gaps. We found this a strong puzzle because it forces participants to treat visualization parameters (e.g., bin size, density bandwidth) as disclosure controls, and it lets us observe how students reason about 
disclosure 
of missingness.
Good solutions emphasize peaks without preserving sample granularity---\eg highlighting the min and max values (Fig.~\ref{fig:gallery}F). 
Constraint violations include plotting the pollution levels as points (Fig.~\ref{fig:gallery}D) or a histogram with 
small bins.

\textbf{3. Show Outliers, Hide Individual Points:}
The sender must communicate about atypical workload or lateness while withholding the identities of warehouses and zones. 
This a strong puzzle because the two data signals are minimally orthogonal: the extremity of a data point is often what makes it identifiable. 
The boundary between constraint adherence and violation is intentionally ambiguous in this puzzle, enabling us to observe how senders decide what degree of specificity is necessary for the receiver to recognize atypical warehouses, and at what point that specificity begins to reveal which warehouses these are. 
Receivers must similarly reason about identifiably, privacy, and acceptable disclosure through 
visualizations.
A good solution summarizes distributions and highlights extremes without plotting raw data---\eg a histogram that reveal tails while preventing re-identification. A clear constraint violation plots each warehouse or zone as a point. 

\subsection{Scoring Solutions}
\label{scoring_solutions}
To define student performance on the game, we develop a 
scoring rubric
that treats show-hide puzzles as constraint satisfaction problems.

\textbf{Rubric Description:} 
We develop a scoring rubric to grade solutions at the level of data signals 
(see Section~\ref{showHide}).
Rather than
treating data signals as simply revealed or 
hidden, we evaluate
constraints formalizing the receiver and sender's disclosure goals as either \textit{satisfied, risked, or broken}.
To support a common evaluation logic for any given data signal, we conceptualize the receiver's information need as a negation of an opposite disclosure constraint---\eg \textit{not hiding} saturation, peaks, or outliers.
For every data signal $s$, the rubric has \textbf{\textit{three components}}: 
(i) $relevantFields$ enumerates the variables needed to instantiate that signal (\eg receiver’s information need or sender’s disclosure constraint), (ii) $markset$ captures which Altair mark types could, in principle, encode those fields in a way that reveals the signal, and (iii) mark-specific
\textit{heuristics} provide criteria for determining signal presence from the rendered visualization, 
incorporating suggested cues for interpreting visual elements and underlying data transformations. 
The rubric for each signal is of the following general form:
\begin{description}[leftmargin=0pt]
  \item[\textsc{isDataSignal}:] It is the case that a mark 
  $m \in \mathit{markset}$ encodes a variable 
  $\mathit{var} \in \mathit{relevantFields}$ such that 
  $\mathit{heuristic}(m)$ indicates the visualization provides sufficient
  evidence to detect data signal $s$.
\end{description}
Appendix Table~\ref{tab:data-signals-scoring-rubric} lists each data signal included in our puzzles together with the corresponding heuristic rubric used to determine whether a given visualization reveals that signal. 
Figures~\ref{fig:teaser} and~\ref{fig:gallery} demonstrate the application of the rubric to student solutions.

\textbf{Development Process:}
We required a systematic way to evaluate whether a visualization succeeded at revealing or hiding specific data signals.
This was non-trivial because (i) there were many ways to adhere to or violate the same constraint, and (ii) 
success
depended on design decisions made upstream of the ultimate visual artifact, which can be difficult to detect or reason about from the visualization alone~\cite{mehta2025designingDisclosure}.
Developing a scoring system enabled us to analyze visualizations in terms of what data signals they made it possible to infer. 
By centering data signals, we sought to make the scoring system extensible to puzzles beyond those tested in our study (see Section \ref{extending_the_game}), providing a general account of performance on data disclosure tasks.

Before running the game, we explored whether scoring could be automated using formal logic, similar to the definition of \textsc{isDataSignal} above.
We attempted to implement the constraint satisfaction logic as an extension of Draco~\cite{moritz2019-Draco}, which was developed to recommend visualizations based on hard rules and soft design constraints.
Although we posited that checking data signal disclosure was deeply aligned with Draco's computational representation, we encountered two challenges with this approach. 
First, disclosure depended on all data processing operations used to create a visualization~\cite{mehta2025designingDisclosure}, not just the data transformations in the chart specification.
Handling cases of upstream data processing would have required extending Draco to ingest Pandas code, which was outside the scope of our work on \textit{Purrsuasion}.
Second, adherence to disclosure constraints depended on subtle interactions between a visualization's markset and properties of a dataset---\eg a histogram could show or hide data signals depending on its bin size~\cite{correll2019-LooksGoodToMe}.
Scoring such cases required runtime evaluation rather than rule-based program analysis, making a constraint solving approach brittle without brute force pre-computation of realized signal disclosures.

After running the game, we found that the scoring of constraint satisfaction was better conceptualized as a \textit{sociotechnical judgment}. 
Although fully automated checking of constraint satisfaction was possible in theory~\cite{mehta2025designingDisclosure}, in practice, student solutions and in-game communication highlighted the ways in which these judgments were situated (see Section~\ref{results}). 
For example, in Figure~\ref{fig:gallery}B, the longitude and latitude of each store were binned to create a heatmap that at first glance hid store identity while showing high saturation. However, 
sufficiently small bins could isolate a single observation, making location inferable.
Conversely, in the \textit{Show Peaks, Hide Gaps} puzzle, similar binning of pollution values could hide gaps at the risk of distorting peaks.
We dubbed these cases \textbf{\textit{risky visualizations}}: 
solutions that neither fully satisfied nor broke a given design constraint.
Risky visualizations tended to be a short edit distance from violating a constraint, inviting the receiver to either infer a hidden signal or misinterpret a distortion of the signal they need. 
Importantly, this notion of risk emerged from social dynamics and relationships around data, not from pure propositional logic.
We designed the rubric using mark-specific heuristics in order to support such contextual judgments of disclosure adherence.

\section{Deployment-Based Study Methodology}

Ethical 
dilemmas presented by 
the need for selective 
data disclosure~\cite{mehta2025designingDisclosure} are  
(i) hard for students to gain practice with, (ii) difficult to teach about, and (iii) tricky to investigate in a controlled setting
because they are inherently situated and consequential. 
In order to create a setting where all three challenges can be addressed in tandem, we deploy \textit{Purrsuasion} in the classroom. 
The game serves as a shared, concrete activity that aligns instructional and research goals,
treating our students as knowledgeable participants with moral agency whose strategies and reflections can help us study how people reason about visualization design and negotiated data sharing subject to disclosure constraints.
\looseness=-1

\subsection{Class Context, Participation, and Consent}
We deployed \textit{Purrsuasion} as an in-class activity for a data visualization course in Fall 2025. The course was a core requirement for the undergraduate data science major at the University of Chicago. To ensure students had adequate preparation, we scheduled the activity late in the quarter, week 8 of 10, so that students had gained fluency with Vega-Altair and key visualization concepts prior to participating. Specifically, they had experience working with data transformations, choosing encodings, layouts, and color scales, as well as building geospatial, uncertainty, and interactive visualizations. This allowed them to focus more directly on learning objectives around responsible data disclosure.

We dedicated two classes to the activity. Before the first class, the instructor shared a pre-recorded lecture
on deceptive visualization highlighting design choices commonly considered “bad” visualization practices\cite{correll2017-blackhat}. The first class included a lecture on the taxonomy of disclosure tactics 
developed by Mehta et al.~\cite{mehta2025designingDisclosure},
providing students with a framework for reasoning about design alternatives and the role of data transformations.
We also showed
a video demonstrating the game interface in an example round played by the authors.
In the second class, students played the game during the whole 80 minute session. 
We randomized the order of rounds across student groups and randomly assigned students to groups of three. Students rotated roles across rounds. 
All groups completed at least two of the three rounds. See Section \ref{showHide} for additional details about the game rounds. 
At the end of the game, the students filled out an exit survey that asked for their reflections and feedback. 
All students received course credit for participating and earned extra credit based on performance. 
See Section~\ref{scoring_solutions} for our rubric and Figure~\ref{fig:gallery} for examples of scored student solutions. 
\looseness=-1

We received IRB approval to analyze students' game play data. At the end of the first lecture, students were invited to consent to share their game-play data for research. We emphasized that participation was completely voluntary (i.e., a form of data donation) and would not affect their course grades. Following IRB guidelines, teaching staff then left the classroom while a non-instructor author distributed and collected paper consent forms, ensuring that the teaching staff had no knowledge of which students chose to participate. Game play data was collected from all students for grading, but our analysis only includes those who consented (27 of 33 students enrolled). The data was anonymized by the non-instructor author, and the teaching staff accessed the data of consenting students only after final grades were submitted.
\looseness=-1

\subsection{Mixed Methods Analysis}
\label{analysis_method}
We conducted a mixed-methods analysis combining qualitative analysis of gameplay artifacts with quantitative analysis of log data captured by \textit{Cat Plot}’s activity tracking (see Section~\ref{interface}). Gameplay artifacts included sender-receiver conversations, visualizations exchanged, round 
outcomes, and receivers’ written justifications for choosing a round winner. Log data consisted of the Python code executed in each notebook cell by senders throughout a round.
These data sources offered complementary views of students’ disclosure choices. 
We analyzed both in parallel to connect observed communication and decision-making with the low-level steps of visualization design and implementation. 

For the qualitative analysis, two authors conducted open coding of each sender-receiver interaction, identifying episodes relevant to our research questions and recording analytic memos to preserve context. We used the taxonomy of disclosure tactics developed by Mehta et al.~\cite{mehta2025designingDisclosure} as deductive codes to analyze 
the use of data transformations in visualization design.
The following \textbf{\textit{disclosure tactics}} appeared in our analysis:
\looseness=-1
\begin{itemize}[noitemsep]
    \item \textbf{Encoded values:} Choosing what variables get represented.
    \item \textbf{Aggregation:} Using a statistic (\eg count) to summarize. 
    \item \textbf{Banding:} Using cutpoints (\eg quantiles) to derive an interval.
    \item \textbf{Classification:} Partitioning a continuous variable into bins.
    \item \textbf{Derived values:} Combining two or more variables into one.
    \item \textbf{Subsampling:} Selecting a subset of records to show.
    \item \textbf{Smoothing:} Interpolating density with a function (\eg KDE).
\end{itemize}

See Mehta et al.~\cite{mehta2025designingDisclosure} for the full taxonomy and Section~\ref{discussion} for reflections on why \textit{Purrsuasion} only elicits a subset of tactics.
We met regularly to calibrate interpretations and maintain consistency in qualitative coding. This iterative process yielded an inductive set of themes describing student interactions during gameplay. We used these themes to characterize how students negotiated disclosure, articulated information needs, and reasoned about communication, problem solving, and ethics in visualization design. 

For the quantitative component, we summarized log traces to characterize senders’ analysis and authoring behavior (\eg design fixation vs. exploration) and used these summaries to corroborate and enrich the qualitative findings. See Supplemental Materials for the students’ gameplay data and a codebook containing the full analysis.

\section{Results}
\label{results}
\begin{figure*}[!t]
    \centering
    \includegraphics[width=\textwidth]{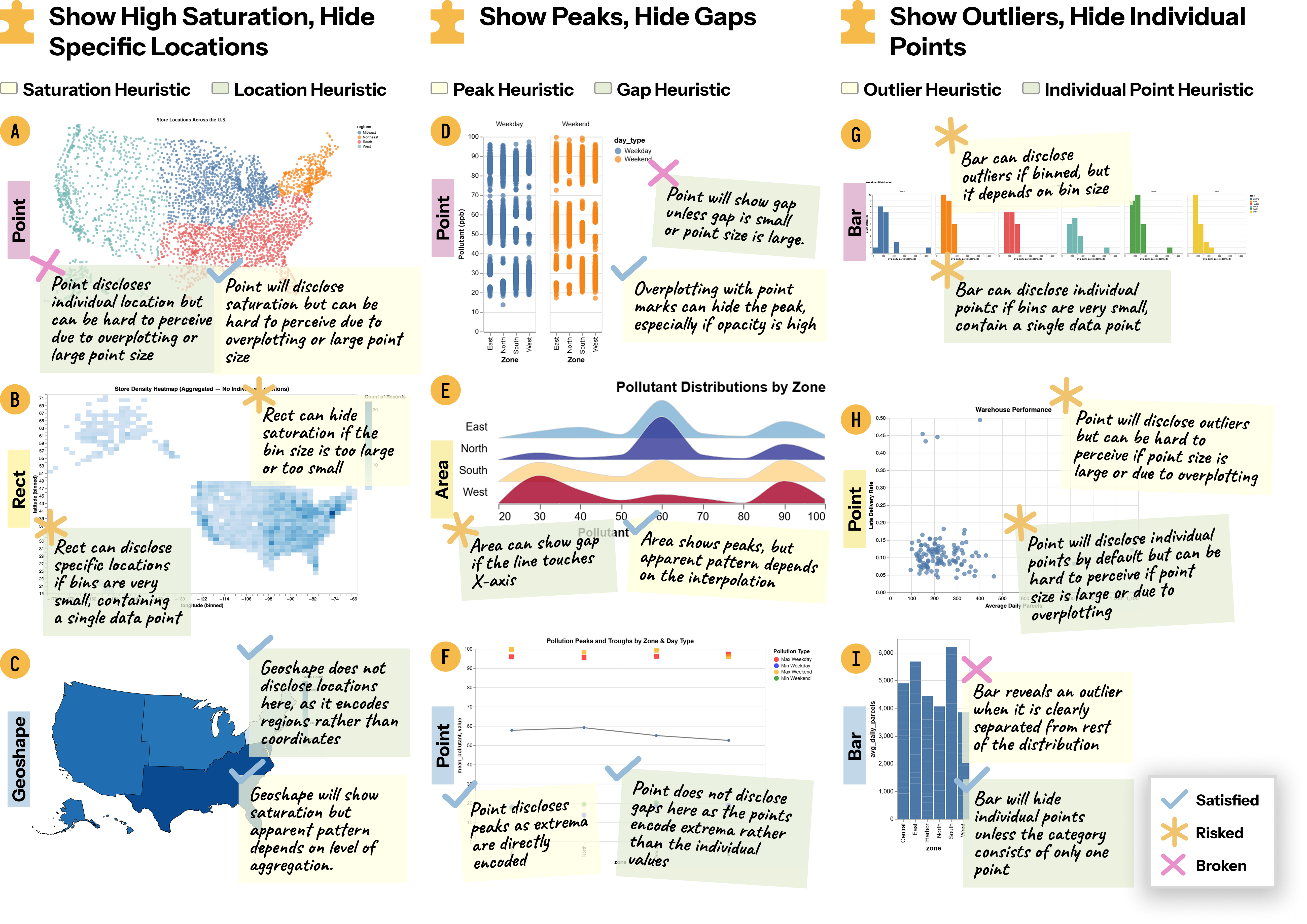}
    \caption{
   Example student-authored visualizations across all three puzzles in our deployment. Each visualization is annotated with the heuristics used by instructors to assess whether the solution reveals the data signal for information need and hides the data signal for the disclosure constraint.
    \label{fig:gallery}
    }
    \vspace{-2em}
\end{figure*}

We found that students negotiating data disclosure confronted a \textit{\textbf{"gulf of envisioning" in visualization}}\cite{subramonyam2024_gulfofEnvision}: it was difficult for students, both as senders and receivers, to conceive of an optimal solution to the show-hide puzzles we tested.
This drove mirrored themes that our analysis uncovered in sender and receiver behavior, respectively.
For senders, difficulty imagining what an optimal solution would look like precipitated \textit{satisficing in visualization authoring}, where students exhibited design fixation~\cite{parsons2021_designFix, jansson1991_designFixation} 
and gulfs of execution around designing for disclosure with grammar of graphics APIs~\cite{Wu2024-transforms,mehta2025designingDisclosure}.
For receivers, the same difficulty imagining an optimal solution precipitated an \textit{inability to attribute authorial intent} on the part of senders.
Given the information asymmetry between senders and receivers and the invisibility of the sender's labor in visualization authoring, receivers tended to judge the work of senders in terms of the utility of the visualizations they sent for the task, rather than by confidence in the sender as a moral actor.
Our analysis traced these themes across students' communication, problem solving, and ethical reasoning.

\subsection{Communication in Sender-Receiver Triads}

We analyzed student communication to characterize (1) how receivers operationalized ambiguous data needs, (2) how senders signaled competence and trustworthiness through interaction, and (3) how interface constraints like limited exchanges shaped negotiation dynamics. 

\textbf{Receivers differed in how they first requested data.} 
Across 25 receiver turns, most receivers (15/25) began the conversation by asking for task-specific information about the data, often borrowing language from the instructions we provided to them. A smaller subset (7/25) started by probing the dataset more broadly—asking what patterns or variables were present—before narrowing to a task-specific follow-up. For example, receiver \pxx{31} asked the senders, \textit{``Please send data on warehouses so I can see any unusual patterns,''} and then pivoted in the second message to a more explicit outlier-focused request. Only two receivers (\pxx{16},\pxx{18}) took a different approach by requesting for a general overview of the data before asking for specific task-related information. Taken together, these openings reflected different ways receivers approached early uncertainty about how to what evidence was available in the dataset and how to allocate a limited number of queries.

\textbf{Second-turn receiver requests fell into a small set of recurring request types.} In the 19 rounds where receivers sent a follow-up, their second message typically served one of five purposes. Most commonly (7/19 responses), receivers requested new evidence beyond what a sender had shown. This included asking for additional variables or alternative ways to surface the target signal---\eg \textit{“Are there any other variables (besides zone) that can be added to find outliers…?”} (\pxx{31}). This extended to puzzle-specific needs for additional evidence: in the \textit{Show High Saturation, Hide Specific Locations} puzzle, 4/8 receivers (\pxx{13}, \pxx{20}, \pxx{22}, \pxx{26}) asked for population context, suggesting raw counts alone were hard to interpret for saturation.
4/19 follow-ups exhibited clarifications about how to interpret a visualization. For instance, \pxx{12} asked one of the senders, \textit{``I don’t really understand this graph… So you have data for east, north, south and west?''} Less frequently, receivers asked for alternative visualization design choices to better support their 
task---\eg \textit{"…more granular detail… state by state… and try to make a more diverging color scale”} (\pxx{6}). Finally, some responses (2/19 responses) requested statistical information to validate that the dataset supported their inference like \pxx{16}, \textit{“…show some summary statistics… so I can be sure there are outlier warehouses?”} while others (2/19 responses) like \pxx{20} requested for a specific visualization type, \textit{“Can you show me if there are any outliers by zone or if there is a smaller region? Could use a boxplot.”} 
 
\textbf{Receivers almost never integrated evidence across senders.} Although each receiver interacted with two competing senders who had access to the same dataset, receivers often responded to the first visualization they saw rather than waiting to compare both senders’ visualizations. This produced a first-response anchoring dynamic: the second turn was typically used to clarify or extend the first visualization received, foreclosing the opportunity to triangulate evidence across sellers or request complementary views.
We expected more strategic behavior, especially when both senders responded at roughly the same time. It suggested that receivers prioritized making a single view interpretable and actionable for each sender.
We also observed that the puzzle structure shaped the feasibility of cross-sender integration such as in the \textit{Show High Saturation, Hide Specific Locations} puzzle, where sender--receiver interaction was typically brief, leaving little opportunity for iteration or for senders to provide complementary views at multiple aggregation levels. 
Across 25 rounds, we found only one instance of an explicit cross-sender reference. In this case, after seeing sender \pxx{16}’s visualization, receiver \pxx{29} used it to formulate a more targeted request to sender \pxx{6}, asking about a dataset variable that had not been visualized in \pxx{6}’s initial response.

\textbf{Sender text functioned primarily as narration and alignment rather than reasoning.}
Across 66 visualizations, over half (37/66) were accompanied by a message that largely restated what the chart depicted. 
13/66 were generic handoffs such as, \textit{“Please see attached,”} (\pxx{6}). Text more rarely functioned as coordination: in 9 responses, senders explicitly checked alignment with the receiver’s intent, \eg \textit{“Does this work?”} (\pxx{19}). When senders expressed constraints on what they can show, 
only two responses explicitly cited the disclosure constraint. More commonly senders attributed omissions to dataset scope or interface friction such as, \textit{“Population isn’t included”,} (\pxx{32}), or \textit{“We only have weekday vs weekend,”} (\pxx{6}), making it difficult for receivers to distinguish structural limitations from strategic withholding motivated by disclosure constraints.
Senders seldom provided dataset-level context. Only one sender (\pxx{18}) included a brief schema description. 
Additionally, senders rarely articulated the reasoning behind their design decisions, such as how they determined which points counted as outliers. This ambiguity made authorial intent harder to attribute and trustworthiness harder to judge, leading receivers to seek clarification.


\subsection{Sender Problem Solving}
\label{student_prob_soln}

Our analysis surfaced evidence of a ``gulf of envisioning''~\cite{subramonyam2024_gulfofEnvision} in senders' visualization authoring. 
Students exhibited design fixation that we interpreted as a struggle to ideate optimal solutions to meet disclosure goals.
Further, students underutilized data transformations (i.e., disclosure tactics~\cite{mehta2025designingDisclosure}, see Section~\ref{analysis_method}) as an approach to eliminating the signal subject to the disclosure constraint in each puzzle.

\textbf{Senders often approached disclosure through a \textit{vagueness-first} strategy.} 
They began with coarse or minimal evidence and increased specificity only when asked. 
As \pxx{6} explained, \textit{"I basically tried to create graphs that were as vague as possible, and when prompted for more would only provide exactly what the person asked for."} 
Students achieved this primarily through the \textit{encoded values} disclosure tactic, which they used in all 66 exchanged visualizations to limit what variables were encoded.
We found this a natural strategy in response to the disclosure constraints in \textit{Purrsuasion}, all of which 
designated
protected fields such as \texttt{zone} in the \textit{Show Outliers, Hide Individual Points} puzzle.

\textbf{Senders overused aggregation as a disclosure tactic.}
45 out of 66 visualizations used \textit{aggregation} to collapse data values into summary statistics, with 13 of these also using \textit{classification} to bin values before summarizing.
In our log data, aggregation accounted for 73.7\% of unique executions containing data transformations. 
As \pxx{22} described, \textit{"I focused on aggregated views like region or broader categories, basically just providing summaries, so the receiver could still see trends without being able to pinpoint exact sources."} 
Students may have gravitated toward aggregation due to ease of implementation and familiarity---\eg bridging the gulf of envisioning with a disclosure tactic whose influence on available signals they could readily anticipate.
However, using aggregation sometimes yielded risky visualizations that only partially revealed the receiver’s information need.
For example, in Figure~\ref{fig:gallery}I, aggregating \texttt{avg\_daily\_parcels} by \texttt{zone} protected the identity of individual warehouses but distorted the receiver's ability to detect outliers.
Interestingly, 22 out of 25 round winners used aggregation even though this produced risky visualizations in 11 out of 22 cases (Fig.~\ref{fig:solution-matrix}).
In contrast with the overuse of aggregation, we found that other disclosure tactics were underutilized even in scenarios where they produced an optimal solution (\eg \textit{banding} in Fig.~\ref{fig:gallery}F).

\begin{figure}[t]
    \centering
    \includegraphics[width=\columnwidth]{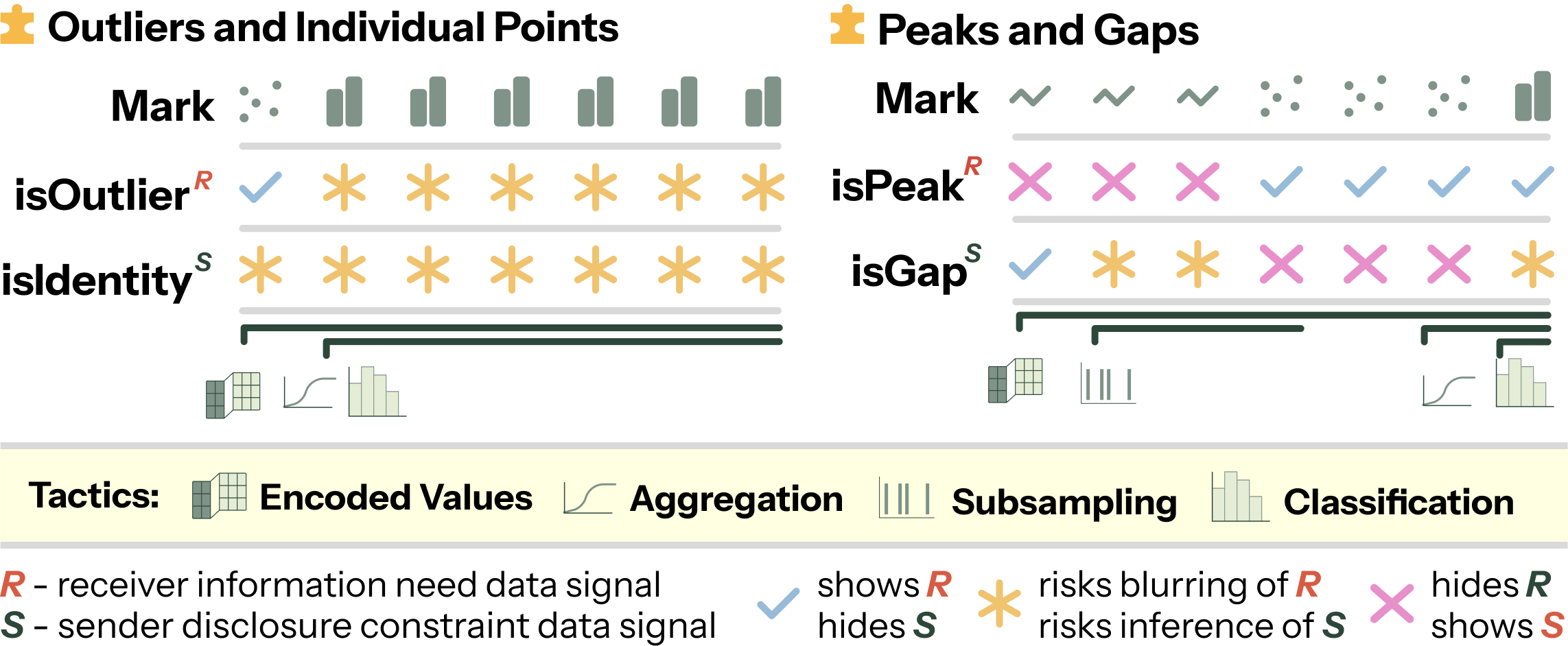}
    \caption{
    \pxx{33}'s chart creation sequences up to their sent visualizations. 
    }
    \label{fig:anchor_log_data}
\end{figure}

\textbf{Senders \textit{anchored on ``safe'' chart types}.}
In their exit surveys, many senders described selecting a chart type early on in their design process and then adjusting encodings to satisfy the disclosure requirements, rather than beginning with data transformations that reduce inferential access and then choosing a chart to match. 
For example, \pxx{33} describes, \textit{"I first chose a type of graph (line chart, histogram, choropleth, etc.) and then carefully thought about the encodings to make sure I satisfied the requirements. If the visualization violated my constraints, then I'd start again with this process."} 
\textit{Purrsuasion}'s log data corroborated this chart-anchoring behavior broadly: for 16 out of 27 students, the first chart type they rendered became an anchor. Across puzzles, the anchor chart type on average accounted for 63.1\% of a student's total visualizations iterations. 
However, the chart that students perceived as ``safe'' was not always so.
For example, Figure~\ref{fig:anchor_log_data} summarized the sequences of visualizations that \pxx{33} authored across two puzzles, showing that in the \textit{Outliers and Individual Points} puzzle, they exhibited design fixation on a risky histogram.
In contrast, Figure~\ref{fig:anchor_log_data} showed that \pxx{33} struggled to ideate a solution that didn't break constraints on the \textit{Peaks and Gaps} puzzle, suggesting that defaulting to a ``safe'' visualization may have reflected satisficing due to the ``gulf of envisioning''~\cite{subramonyam2024_gulfofEnvision} more so than student recognition of promising solutions.

\textbf{Students exhibited uncertainty about what counted as violating a disclosure constraint.}
We found recurrent breakdowns in how senders translated the hiding requirement into an actionable design check.
For example, in Figure~\ref{fig:gallery}D, plotting \texttt{ppb} against \texttt{zone} made discontinuities where data were not recorded visible to the naked eye. 
In other cases, similar confusion led to risky visualizations---\eg in Figure~\ref{fig:gallery}H, by creating a scatterplot of \texttt{avg\_daily\_parcels} against \texttt{pct\_late\_deliveries}, the sender did not directly reveal warehouse identities but provided enough information for an interested receiver to follow up by asking about the protected signal.
We found that such cases were most common in the \textit{Outliers and Individual Points} puzzle, where the information need and disclosure constraint 
were least orthogonal and disclosure adherence was most open to interpretation.
This uncertainty meant that students were not always aware of their performance.
For example, consider \pxx{34}: \textit{"On the first round I had completely missed the constraint, but on the second round as sender, I made sure to check the columns I had available to me thoroughly to and eliminated columns that were unavailable due to the constraints."} In reality, 
\pxx{34} broke constraints in both rounds, illustrating that students' a vagueness-first strategy, which limits \textit{encoded values}, was only as reliable as the sender's ability to envision solutions.

\subsection{How Receivers Selected Contract Winners}

\begin{figure*}[!t]
    \centering
    \includegraphics[width=\textwidth]{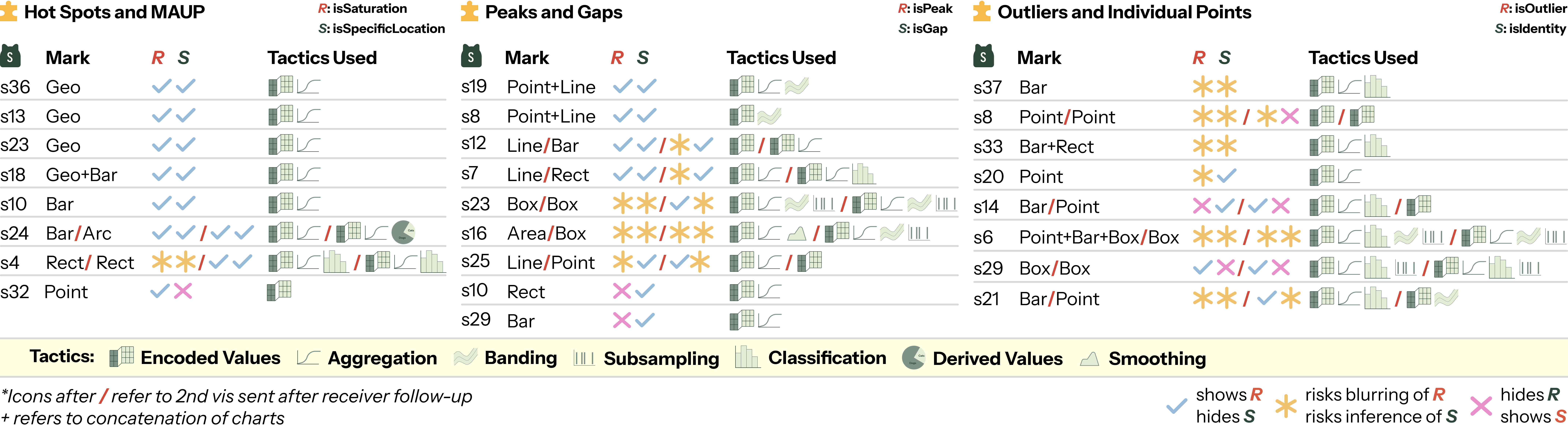}
    \caption{
    For each round winner (grouped by puzzle) we show: participant id, mark type, disclosure adherence per signal, and disclosure tactics used.
    \label{fig:solution-matrix}
    }
    \vspace{-2em}
\end{figure*}

Receivers based their justifications for choosing contract winners on the epistemic utility of visualizations rather than explicit ethical reasoning.
For instance, 23 out of 25 receivers awarded a contract to a sender with visualizations that either satisfied or risked their information need (Fig.~\ref{fig:solution-matrix}). 
Receivers frequently prioritized apparent interpretability. 
For example, 
\pxx{8} preferred a visualization because, \textit{"This data more clearly conveys the information I'm looking for."} 
Color encodings were a popular interpretability cue, with receivers saying things like,
\textit{"Love the color encodings that show clearly which zone of warehouse have outliers in lateness,"} (\pxx{37}), and, 
\textit{"It uses the different colors for each region to visibly show spread!"} (\pxx{24}).

Some receivers relied on 
judgments of whether visualized data aligned with their expectations about the domain,
consistent with prior research~\cite{kim2018_dataThroughOthersEyes}. 
This was illustrated by \pxx{6}’s rationale: \textit{"Although the other graph has a state-by-state stratification, I think this one is better just because the numbers make more sense. The biggest thing throwing me off of the other map says that there are 250 stores in Alaska which absolutely makes no sense."} It is important to acknowledge that the student was correctly detecting an artifact of our synthetic data generation process, as we seeded this puzzle’s data using an airports dataset, producing an unusually high count in Alaska. This observation illustrates a challenge in designing realistic puzzle datasets
(see Section~\ref{extending_the_game}).


Receivers valued evidence that more directly supported the task and provided sufficient data coverage. 
For instance, \pxx{36} explained that \textit{"\pxx{7} provided cleaner, more targeted visualizations that directly addressed the goal of identifying peaks and troughs, and adapted quickly when asked for a different view."} 
\pxx{21} emphasized, 
\textit{"This visualization actually provided information about late deliveries, whereas the first visualization only really gave a summary about daily parcels."}
Others similarly highlighted task fit, informativeness, and adaptability across senders.


\subsection{Student Reasoning about Ethical Data Communication}


Students reasoned about the ethics of data communication across multiple aspects of the game, including the design choices involved in making visualizations, the competing incentives of senders and receivers, and the demands of disclosure constraints.
For example, \pxx{34} reflected,
\textit{"I think it changed my thinking because of how easy it is to try to lie and/or fabricate data to meet the needs of the receiver. You have to think carefully about the choices you make, as it is easy to make visualizations that mislead the receiver into thinking you have what they need."}
Retrospectives like this highlighted how the game made ethical dilemmas salient and actionable.
Occasionally, senders directly stated their constraints as way of resolving the central dilemma presented by the need for selective disclosure---\eg \textit{“Can't show you exact state data but you can make out general trends from this,”} (\pxx{16}).
The tension between sender-receiver incentives also emerged 
more fully
once students had the opportunity to play both roles. 
As \pxx{8} described, \textit{"As the sender, it was tough to adequately communicate what the receiver wanted while still maintaining my ethical standards. Conversely, as the receiver I ultimately went for the sender who gave me the most info---my standards of ethics (maximal disclosure) might not have lined up with their specific goals."}
Rather than treating full disclosure as a fixed principle, 
this student recognized
it as a negotiated tension: what feels desirable from the receiver’s side is not always what feels appropriate or permissible from the sender’s side.


\section{Discussion}
\label{discussion}

By developing and deploying \textit{Purrsuasion}, we advance methodology and knowledge around visualization authoring and interpretation in scenarios that involve negotiated data disclosure.
The game platform itself provides observability into how participants navigate visualization design and interpersonal correspondence around data sharing when balancing simulated tensions between responsibilities to different parties.
Our codification of disclosure problems in terms of data signals, show-hide puzzles, and a heuristic rubric for evaluating candidate designs (see Section~\ref{the_game}) lays the foundation for new ways of studying visualization (see Section~\ref{extending_the_game}).
Based on our deployment of \textit{Purrsuasion} in a visualization class for data science undergraduates at the University of Chicago, we draw out broader implications of our study for trust, design ideation, query formulation, and visualization pedagogy.

\textbf{Receivers face an intent attribution gap when judging visualizations.}
In most rounds of gameplay, receivers assess 
the utility of a visualization for meeting their information need
more readily than they
assess the intentions, constraints, or design process of the senders who created it. 
Accordingly the utility of a visualization serves as a proxy for the trustworthiness of the sender.
Although receivers in \textit{Purrsuasion} could in principle triangulate claims across sender responses and treat disagreement as a signal for further scrutiny, we only observe one instance of this strategy.
Future work should explore ways of helping audiences interpret visualizations more defensively, \eg by enabling them to reason proactively about (i) what a visualization might not show them by design or (ii) how difficult it would be to verify a suspected signal in a risky visualization.
More broadly, visualization tools should support trust
formation 
not only through
surface cues like
readability, but through 
analysis of the design process behind an image.

\textbf{Disclosure constraints lead senders to satisfice rather than explore designs.}
Senders 
tend
to settle on the first 
visualization they find that satisfies their disclosure constraint,
rather than exploring a broader range of possible solutions. 
While disclosure constraints 
narrow the available design space, they do not by themselves explain the degree of design fixation we observe. 
Our findings instead suggest that a central difficulty is ideational: senders struggle to envision multiple acceptable ways of satisfying both the receiver's 
information need
and the disclosure constraint at once.
Although this may be partially attributable to the 20-minute round structure in our deployment of \textit{Purrsuasion}, it also suggests opportunities for future work on authoring interfaces supporting rapid ideation under disclosure constraints (cf. recent AI-assisted visualization tools\cite{Maddigan2023_Chat2VIS, dibia2023-lida, Tian2025-ChartGPT}).

\textbf{Receivers struggle to articulate information need.}
We find that receivers mostly treat written requests for information as narrow queries for relevant evidence, but lacking initial insight into available data, they seldom formulate these queries with a beneficial level of precision.
Instead, receivers make vague references to the data signal named in their instructions.
They rarely use their first turn to request a summary that might inform a strong follow-up question, in contrast to the ``overview first''~\cite{shneiderman2003eyes} opening move in most visual data exploration workflows.
Future work should investigate how people formulate queries in negotiated data sharing when
communication timing, interface conditions, and task framing vary.
More broadly, visualization research should not assume that data seekers will spontaneously ask information-optimal questions and 
instead develop explicit support for query formulation.

\textbf{Risky visualization is a dyadic problem.}
Our findings show that negotiated data disclosure produces visualizations that cannot be classified as 
simply ethical or deceptive. 
Rather, visualizations can create data communication risks that range from benign to serious depending on sociotechnical factors such as the data context and the relationships~\cite{Akbaba2025entanglements} of different stakeholders.
For this reason, we argue that work on
visualization 
authoring, literacy, and pedagogy
should expand from
a focus on
individual chart construction and interpretation toward 
an emphasis on
the mutuality of how designers and audiences reason together 
given
asymmetric access to data. 
In such contexts, sharing a risky visualization should be understood as a valid \textit{epistemic action} that helps to coordinate knowledge between designer and audience.
We argue that the resulting risks, \eg of disclosing sensitive information or miscommunicating, are not mere failure modes but should be studied in terms of their social consequences and opportunities for repair.

\subsection{Beyond \textit{Purrsuasion}: Extending the Game}
\label{extending_the_game}

\textit{Purrsuasion} represents one instantiation of a broader class of \textbf{\textit{disclosure games}}: hypothetical settings where sender(s) and receiver(s) communicate about target data signals using accessible data and provided tools.
This definition of a disclosure game induces a design space of scenarios that can support visualization research and pedagogy, beyond \textit{Purrsuasion}'s show-hide puzzles in a data marketplace setting.
We describe and generalize the \textbf{\textit{core ingredients of disclosure games}}: 
\begin{itemize}[noitemsep]
    \item \textbf{Game Setting} situates the game in a hypothetical context where data disclosure is consequential. Alternatives to \textit{Purrsuasion's} data marketplace setting include intelligence analysts respecting different levels of security clearance, or biostatisticians preserving patient privacy. Games model real-world settings but cannot truly represent high-stakes scenarios with ethical incentives.
    \item \textbf{Roles} define each player’s position in the game including what information they have, what goals they pursue, and what actions they can take. Disclosure games involve senders and receivers, both of which can vary in number and relationship. The role-play involved in disclosure games assigns quasi-personas, but instructions cannot endow the player with another's experiences.
    \item \textbf{Target Signals} describe the relationships among the data signals to be shown/hidden. In \textit{Purrsuasion}, we use puzzles where one signal is shown and another is hidden, but other objectives are possible---\eg showing two or more signals in the same visualization. Puzzles need not have a hide constraint to be considered disclosure games~\cite{mehta2025designingDisclosure}. Any disclosure game that defines signals consistently with \textit{Purrsuasion} can apply or extend our rubric.
    \item \textbf{Data} are the raw information that players can access, visualize, exchange, and evaluate. Disclosure games can vary in both data access and data quality---\eg senders may access the same dataset or work from datasets that differ in completeness or richness. Any disclosure game needs a way to generate data, whether it's synthetic or based on real data. Future work should generalize such data generation to support visualization research writ large. 
    \item \textbf{Communication Structures} determine who communicates with who, who is able to see whose responses, and who has full data access during the round. For example, senders may respond independently, be allowed to communicate with one another, or see each other's messages. Likewise, receivers may interact with senders through private or public channels. Communication structures include the interfaces used for authoring and exchange---\eg AI-assisted authoring or video-based communication.
\end{itemize}

Below, we sketch three extensions of the \textit{Purrsuasion} platform that demonstrate the breadth of the design space of disclosure games and their utility for research and teaching. 
Because the game ingredients change in tandem to represent new design scenarios, each example varies most if not all ingredients.
In each case, we highlight the most important changes relative to our original deployment.

\textbf{Forecasting for a Decision-Maker:} 
Imagine a game where the receiver needs to make a decision, \eg about where to allocate limited resources, and the senders provide public forecasts with uncertainty from predictive modeling to support this choice.
Such decision-making scenarios are common in policy settings such as disaster relief.
The core dilemma is for the senders to author statistical graphics that preserve patterns in the data that are relevant to the receiver's decision.
Relative to \textit{Purrsuasion}, this game's largest changes concern the \textit{target signals} (i.e., forecasts with uncertainty) and \textit{communication structure} (i.e., forecasts are shared in a public forum).
Deploying this game would enable visualization researchers to study how players use sampling- and modeling-based disclosure tactics~\cite{mehta2025designingDisclosure}, which are more common in statistical analysis and uncertainty visualization than in data sharing scenarios like those modeled in \textit{Purrsuasion}.

\textbf{Medical Risk Communication:} 
Imagine a game where the sender is a doctor using historical data to communicate with a diabetic patient (i.e., receiver) about the risks associated with different insulin delivery plans.
Ideally, the sender and receiver collaborate to develop a personalized care plan---\eg the patient asks questions, expresses concerns, and shares personal routines or preferences, thus shaping which visualizations are useful and how they should be explained. 
The core dilemma is for the sender and receiver to mutually make sense of multiple data sources from different perspectives.
Relative to \textit{Purrsuasion}, this game's largest changes concern the \textit{roles} (i.e., one-to-one information sharing) and \textit{data} (i.e., doctor's public data, patient's private information).
Deploying this game would enable visualization researchers to study collaborative trust-building and how evidence is adapted to personal context over time to support decision-making~\cite{subramonyam2017dataDialog}.
As an exercise in perspective-taking that invites reflection on design scenarios, games like this one that explore power dynamics present an opportunity to critique and revise problem framings in visualization~\cite{Parsons2025-BeyondProbSolving}.
\looseness=-1

\textbf{Data Fusion Under Varying Quality:} 
Imagine a game where the receiver already has some data they've used to train a ML model, but they need to supplement this with additional training data from one of the senders. 
The senders have different datasets, with some holding richer data for the receiver's modeling objective than others. 
Such data integration scenarios are common in industry settings such as risk analysis.
The core dilemma for the receiver is to find the sender whose dataset is best for their task, while the senders aim to be chosen regardless of their data quality.
Relative to \textit{Purrsuasion}, this game's largest changes concern the \textit{target signals} (i.e., receiver needs to supplement their training data) and \textit{data} (i.e., senders have different datasets).
Deploying this game would enable visualization researchers to study how players communicate about data integration and comparative data quality.
It also provides a version of the game where receiver performance can be scored normatively.


\subsection{\textit{Purrsuasion} as a Boundary Object}

Interpretive
data communication games, like \textit{Purrsuasion}, offer distinct methodological benefits for visualization research by affording an examination of shared epistemic uncertainties between students and researchers. Functioning as a boundary object~\cite{star1989_translationadAndBoundaryObj}, the game fosters a connection between research, teaching, and practice. For students, it prompts perspective-taking, allowing them to acknowledge and navigate 
contextualized design
tensions ~\cite{Tanner2022_moralSensitivity}. 
For researchers and educators, the lack of canonical solutions 
or a prescribed
design processes fosters epistemic humility, creating opportunities to learn from  students~\cite{brandt2004_collabThroughGames}. 
For instance, despite our initial intention to score visualization solutions on show-hide puzzles using an automated approach (see Section~\ref{scoring_solutions}), student engagement instead pushes us to characterize risky visualizations as requiring situated interpretation and to develop a heuristic rubric to support sociotechnical judgments of disclosure adherence.


Framing the game as a boundary object also allowed us to explore a richer conception of visualization ethics for both research and pedagogy. \textit{Purrsuasion} demonstrates that integrating ethics into the classroom has the opportunity to cultivate reasoning about morally gray scenarios without coercing students to play out ethically compromising situations. In this regard, we made choices in our puzzle design sensitive to human values and the social contexts of technology use \cite{friedman2013value}. Students act as direct stakeholders negotiating a contract, while constraints over the puzzle's dataset function as proxies for indirect stakeholders who are vulnerable data subjects. For research, modeling and capturing these scenarios through \textit{Purrsuasion} provides a sandbox for observing how values embedded during the design process transform or erode as users translate abstract ideals through technology \cite{ghoshal2023design}.


\subsection{Limitations}
The study was conducted under several important constraints. Because we deployed the game within an 80 minute class session, we had to balance the number of puzzles against the time students had to engage meaningfully with each round. This also led us to limit the receiver to a single follow-up question, so that senders could have time to produce a reasonable visualization. As a result, the interaction patterns we observed reflect a time-bounded form of problem solving. Future work should examine how reasoning about disclosure unfolds in settings where players have more time and no fixed message limits.

By situating the game as a mock data marketplace, we centered puzzle design on concerns such as privacy and data quality. As in any empirical study, the cases we sample shape the phenomena we are able to observe. Our findings thus speak most directly to settings in which data seekers must judge whether limited disclosure provides enough confidence to proceed with a purchase that would later reveal the full dataset. Future work should examine how the game translates to other constrained data communication settings, including cases that require statistical inference, communication of multiple data signals at once, or interpretation by viewers with varying levels of visualization literacy.

The study was conducted in an undergraduate visualization course, so the behaviors we observed reflect a particular participant population. Our participants likely occupied an intermediate position between visualization novices and practitioners, with greater technical fluency and visualization literacy than the average audience, but less experience than data scientists or visualization experts. Future work should investigate how these behaviors vary across populations with different levels of technical expertise, visualization experience, and domain knowledge.
\looseness=-1

Finally, while we allowed students to use generative AI tools during gameplay, we did not have the ability to directly observe their interactions with these systems. Our understanding of how AI influenced problem-solving, idea generation, and coding errors was limited to sparse self-reports and post-hoc analysis of Altair code snippets.
Future experiments should actively study the effects of AI on behavior in disclosure games and visual data communication writ broadly.

\section{Conclusion}
We contribute \textit{Purrsuasion}, an open-source visualization game for studying how students navigate ethical data communication and negotiated data disclosure. 
Our findings show that 
students authoring visualizations
often satisficed because it was difficult to envision 
solutions that simultaneously satisfied 
all design constraints.
Students acting as the audience for these visualizations
often struggled to infer authorial intent 
and instead based trust on the utility of visualizations for their task.
To evaluate student solutions to puzzles presented in the game, we developed a heuristic rubric that supports sociotechnical judgments of disclosure adherence. Together, these contributions position \textit{Purrsuasion} as both a research instrument and a pedagogical tool for studying and teaching ethical visualization under negotiated data disclosure.
\looseness=-1

\bibliographystyle{abbrv-doi-hyperref}

\bibliography{template}
\newpage
\appendix

\section{Appendices}
\label{sec:appendices}

\subsection{Show-Hide Puzzle Prompts and Data Generation}

Designing puzzles at an appropriate difficulty level was highly iterative and manual. The research team first enumerated candidate show-hide signal pairings grounded in common visualization tasks~\cite{amar2005low, correll2019-LooksGoodToMe, kim2018assessing, sarikaya2017scatterplots, shneiderman2003eyes, wilkinson2008scagnostics} (Table \ref{tab:puzzle_selection}), then screened out pairings that were too adversarial or too easy, resulting in the three puzzles used in the study. To situate each puzzle in real world context, we used ChatGPT to draft role-specific scenarios for receivers and senders. We also used ChatGPT to generate synthetic datasets for each puzzle, sometimes relying on real data as seed (e.g. \textit{Show High Saturation, Hide Specific Locations} puzzle), then manually refined the datasets to meet our requirements. Sender and receiver instructions along with dataset schema are shown in Table \ref{tab:puzzle_prompts}. See Supplemental Material for the dataset CSV files and descriptions of role-specific scenarios.

This process highlighted that while show-hide puzzles are defined by data signal pairings, their alignment with roles, datasets, and game setting is necessary to make the game coherent. In practice, generating puzzles that satisfy all of these requirements simultaneously was challenging. Although GenAI was useful for rapidly exploring scenarios and candidate datasets, careful human curation remained necessary to ensure that the relevant data signals were present and that they could be communicated clearly without collapsing the puzzle’s intended tension. In this sense, scaling show-hide puzzles is less a matter of producing more prompts than of building a reusable design framework for instantiating negotiated disclosure problems across datasets and domains.

\subsection{Walkthrough of Scoring}

Figure~\ref{fig:gallery} shows examples of student-authored visualizations and Table \ref{tab:data-signals-scoring-rubric} consists the scoring rubric of data signals for all the three puzzles. To walk through the scoring process, let us look at the \emph{Show Peaks, Hide Gaps} puzzle . When grading for constraint violation, we apply \textsc{IsGAP} and label each solution as \emph{constraint broken}, \emph{constraint risked}, or \emph{constraint satisfied}. When grading instead for how clearly a sender answers the receiver’s information need, we apply the rubric for the corresponding task signal (e.g., \textsc{IsPeak}).

We now walk through the three example solutions. 
Figure~\ref{fig:gallery}D, the visualization encodes \textit{relevantFields} = \{\texttt{pollution\_ppb}\} using a point mark. Under the \textsc{IsGAP} rubric, point marks reveal gaps unless the gap is visually negligible (e.g., due to large point size or heavy overlap). In this submission, individual points are clearly visible and gaps in the distribution remain visually identifiable, so the prohibited signal is present and the constraint is \emph{broken}. This case can be graded with a direct visual check against the rubric conditions.

In Figure~\ref{fig:gallery}E, the visualization encodes \textit{relevantFields} = \{\texttt{pollution\_ppb}\}with an area mark (a smoothed distribution). For area marks, the rubric treats gaps as recoverable when the density drops to (or near) the baseline over a substantial span of the domain. This submission produces an ambiguous boundary case: smoothing can attenuate gaps, but the rendered densities still approach the baseline in ways that could allow a viewer to infer gaps depending on bandwidth and scaling. We therefore grade this submission as \emph{constraint risked}, and the score is determined by applying the area-mark heuristics directly (baseline contact and extent). 

In Figure~\ref{fig:gallery}F, \textit{relevantFields} = \{\texttt{pollution\_ppb}\} are not shown as individual observations. Instead, the chart encodes summaries (minimum, maximum, and mean by zone) using points and a line. Because the underlying distribution is not displayed, gaps in the distribution cannot be read from the visualization, so the \textsc{IsGAP} signal is not revealed and the constraint is \emph{satisfied}. When a submission relies on derived or aggregated values, grading can proceed by confirming the transformation pipeline using gameplay logs. Although the line mark imposes a visual continuity over a nominal variable (zone), this is an expressiveness violation rather than a disclosure violation, and it does not change the \textsc{IsGAP} constraint adherence.

\begin{table*}
\centering
\caption{Puzzle prompts and data schema used in the study.}
\small
\setlength{\tabcolsep}{4pt}
\renewcommand{\arraystretch}{1.15}

\begin{tabularx}{\textwidth}{@{} p{2.5cm} Y Y >{\raggedright\arraybackslash}p{5.5cm} @{}}
\toprule
\textbf{Puzzle Type} & \textbf{Receiver Prompt} & \textbf{Sender Prompt} & \textbf{Data Schema} \\
\midrule

\textbf{Show Outliers, Hide Individual Points} &
You’re a city logistics planner allocating inspection teams for next month. Outliers in either workload or lateness could signal bottlenecks or failing warehouses. You need to identify which warehouses are atypical so you can plan targeted audits. Your task is to choose a data broker (sender) who you trust to provide this information in full upon signing a contract. &
You’re a data broker selling a dataset on warehouse performance. You want to build trust by demonstrating that the dataset contains the information your potential client (the receiver) needs. However, to protect supplier relationships and avoid pinpointing individual warehouses, you should hide warehouse and zone identities in your visualizations. &
\schemacell{
  \item \texttt{warehouse\_id}: warehouse identifier (character)
  \item \texttt{zone}: city zone (categorical)
  \item \texttt{avg\_daily\_parcels}: average parcels/day (integer)
  \item \texttt{pct\_late\_deliveries}: fraction late in \([0,1]\) (numeric)
}
\\

\addlinespace[2pt]

\textbf{Show Peaks, Hide Gaps} &
You’re an environmental analyst helping identify when and where pollution levels peak or drop significantly. Your task is to choose a data broker (sender) who you trust to provide the data your team needs to guide policy decisions concerning pollution sources or behaviors. &
You’re a data broker selling a high-resolution air-quality dataset. You want to build trust by demonstrating that the dataset contains the information your potential client (the receiver) needs. But to protect proprietary collection patterns, you cannot expose the exact location of gaps in the data distribution in your visualizations. &
\schemacell{
  \item \texttt{reading\_id}: reading identifier (integer)
  \item \texttt{zone}: city zone (categorical)
  \item \texttt{day\_type}: weekday or weekend (categorical)
  \item \texttt{pollutant\_ppb}: pollutant concentration in ppb (numeric, multimodal)
}
\\

\addlinespace[2pt]

\textbf{Show High Saturation, Hide Specific Locations} &
You’re an analyst working with a tenants’ rights group. Your job is to explore how retail stores are spread across the U.S. to understand where saturation is high or low. Your task is to choose a data broker (sender) who you trust to provide the data needed to understand these patterns. &
You’re a data broker who maintains a detailed dataset of retail stores, including their locations and counts across different areas. Your job is to share visualizations showing the dataset contains what your client (the receiver) needs. However, you should not show very fine-grained details (e.g., individual blocks or exact storefront locations) that could enable landlords or large chains to identify stores and raise rents in specific areas. &
\schemacell{
  \item \texttt{ID}: location identifier (character)
  \item \texttt{city}: city name (character)
  \item \texttt{STUSPS}: U.S. state postal abbreviation (character)
  \item \texttt{latitude}: latitude in decimal degrees (numeric)
  \item \texttt{longitude}: longitude in decimal degrees (numeric)
  \item \texttt{fips}: unique federal identifier for U.S. states (integer)
  \item \texttt{regions}: U.S. Census region name (categorical)
  \item \texttt{REGIONCE}: U.S. Census region code in \{1,2,3,4\} (integer)
  \item \texttt{county\_name}: county name (character)
  \item \texttt{GEOID}: county identifier
}
\\
\bottomrule
\end{tabularx}

\label{tab:puzzle_prompts}
\end{table*}

\begin{table}
\centering
\caption{List of show-hide puzzles developed by the research team. Puzzles deployed in the class are highlighted in purple.}
\begin{tabular}{c l l}
\toprule
\textbf{Show} & \textbf{Hide} & \textbf{Difficulty} \\
\midrule
Clusters & Outliers & Hard\\
Pr($>$thres) & Identity & Hard\\
\rowcolor{blue!10}
Peaks/Troughs & Gaps & Hard\\
Differences & Subgroups & Hard\\
\rowcolor{blue!10}
Outliers & Identity & Hard\\
Sample composition & Sample size & Easy\\
\rowcolor{blue!10}
Hotspots/Spikes & Identity & Hard \\
Range & Skew & Easy \\
Monotonic & Gaps & Easy \\
Concave & Range & Easy\\
Relationship A & Relationship B & Easy\\
\bottomrule
\end{tabular}
\label{tab:puzzle_selection}
\end{table}

\begin{table}
\centering
\caption{Counts (V1, V2) by group and puzzle. V1 represents the number of visualizations shared by both senders in the first exchange with the receiver. V2 represents the number of visualizations shared by both senders in the second exchange with the receiver. Puzzle entries where no visualizations were shared are shown as --. 
}
\footnotesize
\setlength{\tabcolsep}{3pt}      
\renewcommand{\arraystretch}{0.95} 

\begin{tabular}{l cc cc cc}
\toprule
& \multicolumn{2}{c}{MAUP} & \multicolumn{2}{c}{OAIP} & \multicolumn{2}{c}{PAG} \\
\cmidrule(lr){2-3}\cmidrule(lr){4-5}\cmidrule(lr){6-7}
Group & V1 & V2 & V1 & V2 & V1 & V2 \\
\midrule
G1 & 2 & 2 & 2 & 2 & 2 & 2 \\
G2 & 2 & 0 & 2 & 1 & 2 & 1 \\
G3 & 2 & 1 & 2 & 2 & 1 & 1 \\
G4 & 2 & 0 & 2 & 0 & 2 & 0 \\
G5 & 2 & 0 & 2 & 1 & 2 & 1 \\
G6 & 2 & 1 & -- & -- & 2 & 1 \\
G7 & 1 & 0 & 2 & 2 & 2 & 1 \\
G8 & -- & -- & 2 & 0 & 2 & 1 \\
G9 & 2 & 0 & 2 & 0 & 2 & 1 \\
\bottomrule
\end{tabular}
\label{tab: frequencyTable}
\end{table}

\newcommand{\MarksetList}{\texttt{Arc, Area, Bar, Point, Line, Rect, Tick, Trail}}
\newcommand{\RelevantFieldList}{\texttt{pollutant\_ppb}}

\begin{table*}[t]
\centering
\caption{
Scoring rubric for all data signals present in \textit{Purrsuasion}. The rubric for \textbf{IsIndividualLocation} follows the same as that of \textbf{IsIndividualPoint}.
}
\setlength{\tabcolsep}{5pt}
\renewcommand{\arraystretch}{1.15}
\begin{tabular}{p{0.19\textwidth} p{0.19\textwidth} p{0.19\textwidth} p{0.19\textwidth} p{0.19\textwidth}}

\textbf{IsGap.} It is the case that $m \in \mathit{markset}$ encodes $\mathit{var} \in \mathit{relevantFields}$ such that \textit{count is zero} over a substantial span of $\mathit{var}$'s domain.
&
\textbf{IsPeak.} It is the case that $m \in \mathit{markset}$ encodes $\mathit{var} \in \mathit{relevantFields}$ such that $\exists$ a contiguous region of the domain where var is higher than adjacent regions.
&
\textbf{IsOutlier.} It is the case that $m \in \mathit{markset}$ encodes $\mathit{var} \in \mathit{relevantFields}$ such that \textit{count is nonzero in a small, isolated region separated} from $\mathit{var}$'s majority domain.
&
\textbf{IsSaturation.} It is the case that $m \in \mathit{markset}$ encodes $\mathit{var} \in \mathit{relevantFields}$ such that \textit{relative concentration} varies over $\mathit{var}$'s domain.
&
\textbf{IsIndividualPoint.} It is the case that $m \in \mathit{markset}$ encodes $\mathit{var} \in \mathit{relevantFields}$ such that $\exists$ an observed record whose value is identifiable and distinguishable from other records.
\\

\begin{minipage}[t]{\linewidth}\vspace{0pt}\small\raggedright\sloppy
\textbf{IsGap (}\par
\textit{markset} = \{\MarksetList\}\par
\textit{relevantFields} = \{\RelevantFieldList\}\par
\textit{heuristic} = \{\par\vspace{-2pt}
\begin{itemize}[leftmargin=*, nosep]
  \item \textbf{Arc:} If a category is empty, the visualization can hide it but the legend will still disclose it.
  \item \textbf{Area:} Area can disclose gaps if the line touches the x-axis.
  \item \textbf{Bar:} If \texttt{bin()}, depends on bin size.
  \item \textbf{Point:}  Points disclose gaps but can be hard to perceive if gap is small or point size is large.
  \item \textbf{Line:} Lines disclose gaps only if var is non-empty, ordered, and encoded on the x-axis.
  \item \textbf{Rect:} Rect can disclose gaps for categorical data.
  \item \textbf{Tick:} Ticks disclose gaps but can be hard to perceive if gap is small or point size is large.
  \item \textbf{Trail:} Trail can disclose gaps if var is non-empty, ordered, and encoded on the x-axis.
\end{itemize}
\}
\textbf{)}
\end{minipage}
&
\begin{minipage}[t]{\linewidth}\vspace{0pt}\small\raggedright\sloppy
\textbf{IsPeak (}\par
\textit{markset} = \{\texttt{Area, Bar, Line, Point, Rect, Tick}\}\par
\textit{relevantFields} = \{\RelevantFieldList\}\par
\textit{heuristic} = \{\par\vspace{-2pt}
\begin{itemize}[leftmargin=*, nosep]
  \item \textbf{Area:} Area discloses peaks but it depends on the bandwidth.
  \item \textbf{Bar:} Bar discloses peaks, if \texttt{bin()}, depends on bin size.
  \item \textbf{Line:} Line discloses peaks but depends on the interpolation.
  \item \textbf{Point:} Points disclose peaks but can be hard to perceive due to overplotting.
  \item \textbf{Rect:} If var is encoded as color, depends on color scale.
  \item \textbf{Tick:} Can disclose but can be hard to perceive due to overplotting.
\end{itemize}
\}
\textbf{)}
\end{minipage}
&
\begin{minipage}[t]{\linewidth}\vspace{0pt}\small\raggedright\sloppy
\textbf{IsOutlier (}\par
\textit{markset} = \{\texttt{Area, Bar, Boxplot, Geoshape, Line, Point, Rect, Tick, Trail}\}\par
\textit{relevantFields} = \{\texttt{$avg\_daily\_parcels$, $pct\_late\_deliveries$, $warehouse\_id$}\}\par
\textit{heuristic} = \{\par\vspace{-2pt}
\begin{itemize}[leftmargin=*, nosep]
  \item \textbf{Area:} Area can disclose outliers but depends on the bandwidth.
  \item \textbf{Bar:} A bar mark reveals an outlier when it is clearly separated from the rest of the distribution. If \texttt{bin()}, depends on bin size.
  \item \textbf{Boxplot:} Boxplot discloses outliers by default.
  \item \textbf{Geoshape:} Geoshape discloses outliers if the outlier is at the level of aggregation.
  \item \textbf{Line:} Line can disclose outliers but depends on the interpolation.
  \item \textbf{Point:}  Points will disclose but can be hard to perceive due to overplotting or large point size.
  \item \textbf{Rect:} Rect can disclose outliers outlier when it is clearly separated from the rest of the distribution but depends on bin size.
  \item \textbf{Tick:} Ticks can disclose outliers but can be hard to perceive due to overplotting.
  \item \textbf{Trail:} Trail discloses outliers if var are encoded as position directly, isolated trails can show outliers, encoding var as width can hide them.
\end{itemize}
\}
\textbf{)}
\end{minipage}
&
\begin{minipage}[t]{\linewidth}\vspace{0pt}\small\raggedright\sloppy
\textbf{IsSaturation (}\par
\textit{markset} = \{\texttt{Area, Bar, Geoshape, Line, Point, Rect, Tick, Trail}\}\par
\textit{relevantFields} = \{\texttt{longitude, latitude}\}\par
\textit{heuristic} = \{\par\vspace{-2pt}
\begin{itemize}[leftmargin=*, nosep]
  \item \textbf{Area:} Area discloses saturation but depends on the bandwidth.
  \item \textbf{Bar:} Bar discloses saturation if \texttt{bin()}, depends on bin size.
  \item \textbf{Geoshape:} Geoshape discloses saturation only at the level of aggregation.
  \item \textbf{Line:} Line can disclose saturation but depends on the interpolation.
  \item \textbf{Point:} Points will disclose saturation but can be hard to perceive due to overplotting or large point size
  \item \textbf{Rect:} Rect discloses saturation but depends on bin size.
  \item \textbf{Tick:} Tick can disclose saturation but can be hard to perceive due to overplotting, clearer with transparency or jitter.
  \item \textbf{Trail:} Trail can disclose saturation if var is encoded as position or width.
\end{itemize}
\}
\textbf{)}
\end{minipage}
&
\begin{minipage}[t]{\linewidth}\vspace{0pt}\small\raggedright\sloppy
\textbf{IsIndividualPoint (}\par
\textit{markset} = \{\texttt{Area, Bar, Boxplot, Geoshape, Line, Point, Rect, Tick}\}\par
\textit{relevantFields} = \{\texttt{$warehouse\_id$, zone}\}\par
\textit{heuristic} = \{\par\vspace{-2pt}
\begin{itemize}[leftmargin=*, nosep]
  \item \textbf{Area:} Area usually hides individual points but depends on bandwidth.
  \item \textbf{Bar:} Bar discloses individual points if one bar corresponds to one record, if \texttt{bin()}, depends on bin size.
  \item \textbf{Boxplot:} Boxplot hides individual points by default but discloses if raw points or outlier points are overlaid.
  \item \textbf{Geoshape:} Geoshape can disclose individual points if var is defined as points within a GeoDataFrame using the geoshape mark.
  \item \textbf{Line:} Discloses if raw observations are connected and individual vertices are readable; smoothing or aggregation can hide them.
  \item \textbf{Point:} Points will disclose individual points by default but can be hard to perceive if point size is large or due to overplotting or large point size.
  \item \textbf{Rect:} Rect marks can disclose specific locations if bins are very small, containing a single data point.
  \item \textbf{Tick:} Ticks can disclose individual points if each tick corresponds to a separate record and remains visually separable.
\end{itemize}
\}
\textbf{)}
\end{minipage}
\\

\end{tabular}
\label{tab:data-signals-scoring-rubric}
\end{table*}

\end{document}